# Reduced graphene oxide thin films as ultrabarriers for organic electronics

*Hisato Yamaguchi[1,2]\*, Jimmy Granstrom[3], Wanyi Nie[2], Hossein Sojoudi[3], Takeshi Fujita[4], Damien Voiry[1], Mingwei Chen[4], Gautam Gupta[2], Aditya D. Mohite[2], Samuel Graham[3], Manish Chhowalla[1]\**

[1]Department of Materials Science and Engineering, Rutgers University, 607 Taylor Road, Piscataway, New Jersey 08854, U.S.A.
E-mail: manish1@rci.rutgers.edu

[2]Center for Integrated Nanotechnologies (CINT), Materials Physics and Applications (MPA) Division, Mail Stop: K771, Los Alamos National Laboratory (LANL), P.O. Box 1663, Los Alamos, New Mexico 87545, U.S.A.
E-mail: hisatoy@lanl.gov

[3]Center for Organic Photonics and Electronics and Woodruff School of Mechanical Engineering, Georgia Institute of Technology, Atlanta, Georgia 30332, U.S.A.

[4]WPI Advanced Institute for Materials Research, Tohoku University, 2-1-1 Katahira, Aoba-ku, Sendai, Miyagi 980-8577, Japan

Encapsulation of electronic devices based on organic materials that are prone to degradation even under normal atmospheric conditions with hermetic barriers is crucial for increasing their lifetime. A challenge is to develop "ultrabarriers" that are impermeable, flexible, and preferably transparent. Another important requirement is that they must be compatible with organic electronics fabrication schemes (i.e. must be solution processable, deposited at room temperature and be chemically inert). Here, we report lifetime increase of 1,300 hours for poly(3-hexylthiophene) (P3HT) films encapsulated by uniform and continuous thin (~10 nm) films of reduced graphene oxide (rGO). This level of protection against oxygen and water vapor diffusion is substantially better than conventional polymeric barriers such as Cytop$^{TM}$, which degrades after only 350 hours despite being 400 nm thick. Analysis using atomic force microscopy, x-ray photoelectron spectroscopy and high resolution transmission electron microscopy suggest that the superior oxygen gas/moisture barrier property of rGO is due to the close interlayer distance packing and absence of pinholes within the impermeable sheets. These material properties can be correlated to the enhanced lag time of 500 hours. Our results provide new insight for the design of high performance and solution processable transparent "ultrabarriers" for a wide range of encapsulation applications.

# 1. Introduction

The performance of organic-inorganic multilayer "ultrabarriers" for encapsulation of organic and other environmentally sensitive electronics is limited by unavoidable pinholes in the inorganic layer (typically $AlO_x$ and $SiO_x$) during vacuum deposition.[1-4] Increasing the thickness of the inorganic film can mitigate this limitation but introduces new defects – cracks induced by stress in the films. State of the art barriers consist of repeated stacking of organic-inorganic units – referred to as "dyad" structure to increase the diffusion length of the gas species. However, this is problematic due to the increased fabrication complexity as well as loss of flexibility. Furthermore, elimination of the vacuum deposition step for ultrabarriers would be beneficial for device fabrication methods such as roll-to-roll processing. Therefore it is desirable to fabricate impermeable, flexible, and preferably transparent ultrabarriers *via* solution processing that is free from complexity of the stacked "dyad" structure.

Graphene is an atomically thin sheet of carbon, which possesses attractive material properties, especially as ultrabarrier layers.[5-8] Its closely packed hexagonal atomic structure prevents permeation of even hydrogen,[9] and being atomically thin makes it flexible and optically transparent. Mechanical exfoliation and chemical vapor deposition (CVD) are known routes to obtain less defective graphene but uniform deposition over large areas on arbitrary substrates at low temperatures may limit their use as barriers.[10-12] We have therefore investigated the barrier properties of graphene oxide (GO) obtained from chemical exfoliation of graphite.[13-15] GO consists of atomically thin sheets of graphene that are decorated with oxygen functional groups and has microscopic lateral dimensions.[16] It is well known that in contrast with pristine graphene, GO is highly defective and chemically heterogeneous. It can be made electrically active by evolving oxygen from its structure but these can introduce holes in the basal plane of the GO sheets.[17] Continuous layers of GO and rGO can be deposited using several techniques[18-20] but it is unclear if impermeability can be introduced in such thin films. Here we demonstrate that GO and rGO layers exhibit excellent hermeticity, and the presence of oxygen functional groups attached to both sides of the graphene basal plane can act as atomically thin organic layers of the "dyad" structure. Although some efforts have already been made to use solution processed and transparent GO as hermetic films, reports generally focus on its incorporation in polymer composites, where GO serves merely as impermeable sheets that are randomly oriented and widely spaced.[21-22] These types of structures create permeation paths in the films, limiting the contribution of barrier effects from the GO. Recently, Yang et al.[23] reported significantly enhanced gas barrier performance using well-ordered GO film-polymer bilayer. However, evaluation of intrinsic gas barrier properties of GO and rGO thin films remain unexplored. The elucidation of their gas barrier properties including the oxygen diffusion dynamics should allow the design of high performance and solution processable transparent ultrabarriers based on two-dimensional nanosheets.

## 2. Results and Discussion

We have monitored the degradation of a common electrically active organic material, poly(3-hexylthiophene) (P3HT), to test the barrier properties of GO and rGO. It has been shown that photo-oxidative degradation of P3HT occurs upon exposure to $O_2$ (also moisture) under illumination, resulting in an increase in the optical transparency.[24] Therefore, monitoring the transparency of the P3HT under humid conditions is a simple and accurate method for determining the barrier properties of GO/rGO thin films. Specifically, photolysis (film degradation due to light exposure in an inert atmosphere) of P3HT occurs at a rate that is several orders of magnitude lower than photo-oxidation for the first few thousand hours of light exposure, thus sufficiently long term oxygen barrier test can be performed using this method.[25] The measurements were made in ambient environment with weak illumination intensity. Illumination was kept low to eliminate possible temperature increase of the samples as well as reduction of GO by light (see methods for details).[26] In addition to GO and rGO layers, control measurements were also made on P3HT alone and Cytop$^{TM}$ covered P3HT. Cytop$^{TM}$ was chosen based on its high gas barrier property, which exhibits up to an order of magnitude lower water vapor transmission rate than other polymer barriers, e.g. polyvinylacetate (PVAc) and polymethylmethacrylate (PMMA).[27]

**Figure 1** (a) shows the normalized transmittance of P3HT films *versus* time for different barrier films (after subtracting transmittance of the barrier material). Thin (10 nm) and thick (20 nm) films of GO along with 15 nm rGO films were prepared by vacuum filtration for evaluation. The inset in Figure 1 (a) shows GO thin films on glass substrate (left), and P3HT encapsulated with thin GO film (right) that are optically transparent. GO films in this thickness range are also flexible as demonstrated in our previous study.[28] Entire P3HT film including the edges was fully encapsulated with GO films (Fig. 1 (b)). The summary of degradation as a function of time for each film is plotted in **Figure 2** (a) (black). As expected, fastest degradation was observed for the bare P3HT film. The transmittance of P3HT showed steady increase, which reached 100% after only 144 hours (hrs) (defined as lifetime hereafter). The lifetime after encapsulation with 400 nm thick Cytop$^{TM}$ improved to 493.2 hrs, a factor of 3.4 increase compared to bare P3HT. This degree of protection against oxygen and water vapor diffusion is consistent with value for Cytop$^{TM}$,[29] indicating the validity of the P3HT test for evaluation of oxygen/moisture barrier property. When the P3HT was encapsulated with GO film, however, the lifetime reached 672 hrs with a film thickness of only 10 nm (green line). The capability of GO films to act as an oxygen barrier with thickness of 10 nm is in sharp contrast to polymer chain-based barrier films including Cytop$^{TM}$, which usually require at least 100 nm of thickness to minimize the influence of pinholes.[27] The lifetime can be further improved by simply increasing the film thickness. Increasing the GO film thickness by a factor of 2 (10 nm to 20 nm) leads to an increase in lifetime of up to ~840 hrs (orange line). The oxygen barrier properties improved substantially after annealing the GO films at 150 $^o$C. The lifetime of rGO barriers reached ~1,440 hrs (red line), almost ~800 hrs improvement compared to non-annealed film with the same number of stacked layers, and 1,300 hrs increase compared to 400 nm thick Cytop$^{TM}$ film. The rGO film with this superior gas barrier property had relatively high optical transmittance of 75% at 550 nm.

Moreover, unlike the other tested barrier films, no increase in the transmittance of the P3HT film was observed in the initial 500 hours with rGO barrier films.

To compare the gas barrier properties of the different encapsulation materials, the lifetime was divided by the film thickness to obtain lifetime per unit thickness as summarized in Fig. 2 (a) (blue). The obtained values were 1.2 hrs/nm for Cytop$^{TM}$ and 67.2 hrs/nm for GO thin film, equating to a factor of ~50 improvement. The non-linearity in the GO film properties with thickness can be attributed to the fact that the thicker films contain a substantial amount of wrinkles that introduce structural inhomogeneity (inset of **Figure 3** (a), scale bar shows 2 μm).[18] These wrinkles can be "ironed" out *via* annealing. For rGO film, the lifetime per unit thickness increases substantially up to ~100 hrs/nm, almost double that of GO with same number of stacked layers, and two orders of magnitudes larger than the Cytop$^{TM}$.

To gain insight into the gas diffusion dynamics and the origin of improved barrier property of GO/rGO thin films, further analysis of the P3HT transmittance test results was performed. Specifically we focused on two regions: the transition phase in which steady state gas diffusion has not been reached, resulting in very small or no observable permeation; and the steady-state regime where diffusion has reached equilibrium, giving a constant flux of gas species permeating through the barrier films (supplementary information, Fig. S1).[1] The analysis allows determination of the dominant phase in the permeation process, as well as the contribution of each region, which provides information on the barrier dynamics and possible mechanisms. The flux of steady-state phase can be obtained from the slope of the gas barrier lifetime test as shown by the dashed line in Fig. 1 (b). Analysis of the slopes shows very similar values for all of GO/rGO barrier films with only a slight increase after thermal reduction; $3.66 \times 10^{-5}$ for both thin and thick GO, and $3.10 \times 10^{-5}$ for rGO films (values are for plots with time in seconds (s), Fig. 2 (b), black). This result suggests that the effect of thermal reduction on the diffusivity of the gas through rGO barrier is minimal. More specifically, the nature of permeation paths for rGO remains similar to that of GO (i.e. the diameter and/or the width, and friction between gas species of interests). The slopes for P3HT with Cytop$^{TM}$ encapsulation and without any encapsulation were $1.93 \times 10^{-4}$ and 0.20, respectively, giving higher steady-state permeability compared to GO/rGO. In contrast to the minor change observed for steady-state flux, the transition phase shows substantial difference between GO and rGO (Fig. 2 (b), red). While lag time (also called "break-through" time, which can be extracted by extrapolation of the slope to the time axis as shown in Fig. 1 (b)) was not present (~0 hr) for both controls and 10 nm GO film but it reached 81.2 hrs for 20 nm GO film, and 510.2 hrs for the 15 nm rGO film. Large difference in lag time between the films while exhibiting only minor differences in the diffusivity strongly suggests that extremely long diffusion lengths dominate and play a key role in the permeation process of GO/rGO.

More specifically, lag time is a function of two parameters, diffusivity and diffusion length as indicated in the equation:[1]

Lag time (s) = $l^2$ (cm$^2$) /6D (cm$^2$/s)          (1)

The diffusivities of oxygen obtained from the slopes in the steady-state regime of P3HT transmittance data were found to be $7.6 \times 10^{-8}$ cm$^2$/s for 20 nm GO, and $6.4 \times 10^{-8}$ cm$^2$/s for 15 nm rGO film. For comparison, the diffusivity of Cytop$^{TM}$ is $4 \times 10^{-7}$ cm$^2$/s.[29] It is clear that the diffusivity of GO/rGO films is not substantially lower than that of polymers, indicating that it is not the origin of superior gas barrier properties of these materials. The diffusion length can be extracted from equation (1) using the experimentally obtained diffusivity and lag time values above (Fig. 2 (b)). The calculated diffusion length for rGO is 8.40 mm, which is 230% of that of GO (see supplementary information for details). The results quantitatively support that long diffusion length enabled by stacking of rGO sheets plays a crucial role in its superior gas barrier property.

Examination of the material properties of the GO/rGO films provides further insight into the diffusion barrier mechanism. X-ray photoelectron spectroscopy (XPS) shows that mild annealing of GO at 150 °C leads to a decrease in the oxygen content from 54 at.% to 34 at.%, most of which can be attributed to the removal of water trapped between the sheets.[30-31] Oxygen functional groups such as hydroxyl and carboxyl remained even for rGO, which could affect the degradation process of the encapsulated P3HT films as well as organic electronic devices as those groups attract moisture (supplementary information Fig. S3). The annealing leads to compaction of the structure as shown in Figures 3 (c). This compaction is also confirmed by AFM measurements that show that the film thickness after annealing decreases from 20 nm to 15 nm. In addition, the mild annealing does not introduce substantial pinhole like defects that are typically found in fully reduced GO.[32] Instead our high resolution transmission electron microscopy (HRTEM) results revealed that the initial random oxygen functionalized structure of GO (Figure 3 (a)) is replaced by the presence of small highly crystalline regions shown in Figure 3 (b). The restoration of hexagonal atomic structure in the form of 2 - 3 nm nano-islands due to removal of oxygen have beneficial effects on barrier property. These islands decrease the interlayer spacing by removal of oxygen group "bridges" as schematically illustrated in side (Fig. 3 (c)) and top view (Fig. 3 (d)). Above proposed mechanism rely on an assumption that AFM and HRTEM results are uniformly distributed over the entire GO/rGO films. Effects of inhomogeneous defects/sheet boundary distrubutions must be taken into account for more accurate permeation mechanisms. It is worthwhile noting water vapor permeates freely through GO (but not rGO).[33] The proposed diffusion mechanism is such that permeated water pillars act as a barrier for gas species. When reduced, removal of oxygen functional groups clogs permeation paths decreasing water permeation to 1/100. It is possible that this is also the case in our system. Assuming that this is the case, our results indicate that oxygen may affect the degradation of P3HT over moisture. Despite the fact that water vapor permeates freely through GO films, their barrier effects were clearly present in the P3HT transmittance test results.

The feasibility of utilizing rGO barriers in real working organic devices was investigated using bulk heterojunction (BHJ) organic solar cells. Specifically, we encapsulated conventional ITO/MoO$_3$/P3HT:PCBM/Al organic photovoltaic (OPV) devices (Figure 4 (a), see Experimental Section for details). ITO was on 1 mm thick glass substrate, and the illumination was exposed from bottom of the device through the glass substrate thus optical transmittance of GO and rGO encapsulation films did not affect the device efficiency. The device characteristics were monitored under ambient conditions and under constant illumination, without humidity/moisture control. As shown in current-voltage (I-V) characteristics of the devices for 0 hour and 50 hours after exposure to ambient conditions (Figure 4 (b) and (c), respectively), GO/rGO encapsulation showed clear barrier effect on working organic devices. The absolute PCE for starting devices ranged between 3.2-3.6% (supplementary information Fig. S4). The results show that GO/rGO encapsulation improved the normalized power conversion efficiency (PCE) by 20% over the period of 50 hrs (**Figure 4** (d)) compared to devices without encapsulation. We found that J$_{sc}$ of the OPVs exhibited substantial changes while FF and V$_{oc}$ remained mostly unchanged. Fig. 4 (e) shows the normalized J$_{sc}$ for rGO (red circles) and GO encapsulated BHJ solar cells (blue triangles) compared to control devices (black triangles, without encapsulation). J$_{sc}$ for the device with rGO encapsulation showed less than 5 % decrease after 50 hrs, in contrast to the control devices that show a decrease of nearly 30 %. Interestingly, there were no changes in the J$_{sc}$ values for the first 15 hrs in GO encapsulated devices. This period could be viewed as lag time and it indicates that the lag time of rGO encapsulated devices could be over 50 hrs. The fluctuation of J$_{sc}$ observed for rGO encapsulated devices in the initial few hours of testing can be attributed to degradation of polymer active layer of the device (P3HT:PCBM blend) induced by the environment (exposure to air and strong illumination), which was also observed as a steeper slope in the initial 5 hrs of the non-encapsulated device.[34]

## 3. Conclusion

Solution processed, and optically transparent GO/rGO thin films exhibited improved oxygen/moisture barrier performance over a commercially available Cytop™ polymer film that has an order of magnitude larger thickness. The barrier performance and lifetime of GO was substantially improved simply by annealing the films at low temperature to obtain reduce GO (rGO). Characterization revealed that the close packing of interlayer distance in addition to restoration of pristine graphene nano-islands induced by the annealing cause clogging of permeation paths, leading to long diffusion lengths, hence long lifetime. The long diffusion length of rGO film was evident from the enhanced lag time. Finally, the gas barrier property of GO/rGO thin films was successfully demonstrated on bulk heterojunction (BHJ) solar cells. Our results provide a pathway for realizing high performance and solution processable transparent "ultrabarriers" that are suitable for wide range of encapsulation applications, including organic electronics.

## 4. Experimental Section

*Sample preparations and gas barrier measurement setup:* Graphene oxide (GO) was prepared by modified Hummer's method.[18] Briefly, graphite powders were chemically oxidized, exfoliated, and purified by repeated centrifugation. No sonication was applied throughout the process to eliminate formation of pinholes in the GO sheets. Different amount of GO aqueous solution was vacuum filtrated to prepare uniform films of GO with different thicknesses,[18] which were then deposited directly onto the P3HT films for the gas barrier measurements (refer to Figure 1 (b) for the structure of each samples). No post-deposition transfer of the GO films was necessary. Typical film size was 2.54 cm × 2.54 cm (1 inch × 1 inch), with as synthesized films thickness of 10 and 20 nm for thin and thick films, respectively. Some of the films were thermally annealed at 150 °C for 15 min to induce thermal reduction.

The poly(3-hexylthiophene) (P3HT) transmittance test for gas barrier measurements was performed by using spin-cast P3HT films as oxygen sensors. Organic solvent such as toluene and dichlorobenzene, commonly used for solar cell fabrication was used for dissolving P3HT. The polymer barrier used as a comparison was made by spin-casting a solution of commercially available perfluorinated polymer Cytop™ to have the thicknesses of >400 nm. An extra care was taken to minimize the light exposure and oxidization of P3HT films during all of the sample preparation procedures. Specifically, the whole processes was performed in the dark to minimize the exposure of P3HT to the light, and P3HT films were always kept in the vacuum desiccators under dark when not in use. The light source of 22 W fluorescent lamp with mean lumens of 783 was used for the test unless stated otherwise. The typical distance between the lamp and the sample surface was ~5.08 cm (2 inches), where the average power from the lamp was about 0.008 W/cm$^2$. In order to accelerate the testing process, some of the measurements were made using AM 1.5 solar simulator as a light source with power of 100 W/cm$^2$ at the sample surface. Transmittance of P3HT film encapsulated by barrier films are measured and compared with transmittance of reference P3HT (without barrier film) to obtain change in the transmittance of the P3HT films used as sensors.

*Material characterizations:* Atomic force microscopy (AFM) was performed using Digital Instruments Nanoscope IV in tapping mode with standard cantilevers with spring constant of 40 N/m and tip curvature <10 nm. X-ray photoelectron spectroscopy (XPS) has been performed with a Thermo Scientific K-Alpha spectrometer equipped with Al Kα micro-focused monochromatized source (1486.6 eV) which has an energy resolution of 0.6 eV. Spot size was fixed to 400 µm and the operating pressure was kept at $5 \times 10^{-9}$ Pa. High resolution transmission electron microscopy (HRTEM) was performed using JOEL JEM-2100F TEM/STEM with double spherical aberration (Cs) correctors (CEOS GmbH, Heidelberg, Germany) to attain high contrast images with a point-to-point resolution of 1.4 Å. The lens aberrations were optimized by evaluating the Zemlin tableau of an amorphous carbon. The residual spherical aberration was almost zero (Cs = -0.8±1.2 µm with 95 % certification). The acceleration voltage was set to 120 kV which is the lowest voltage with effective Cs correctors in the system. The region of interests was focused and then recorded with total exposure of less than 20 s (0.5 s exposure time for the image).

*Bulk heterojunction solar cell fabrication and their lifetime tests:* Pre-patterned Indium Tin Oxide coated glass slides were cleaned thoroughly in distilled water, acetone and isopropyl alcohol by sonication bath for 15 min respectively and dried on a hotplate on 120 ºC for 30 min. After exposed in oxygen plasma for 3 min, a solution of poly(3,4-ethylenedioxythiophene) poly(styrenesulfonate) (use as purchased, Clevios Al 4083) was spun-coated on top of the ITO slides at 4000 rpm for 40 s resulting in a 40 nm layer. Once the film was dried at 120 ºC for 30 min, a blend solution containing 15 mg Poly(3-hexylthiophene-2,5-diyl) (P3HT, as received from Sigma-Aldrich) and 12 mg [6,6]-Phenyl C61 butyric acid methyl ester (PCBM) in 1ml chlorobenzene was spun-coated at 1500 rpm for 35 s forming approximately 150 nm active layer in argon filled glovebox. Finally, the devices were transferred to vacuum chamber pumped down to $1\times10^{-7}$ torr for aluminum deposition. The device measurement was carried out in air without humidity/moisture control (relative humidity (RH) of 50-60 %), using 100 mW/cm$^2$ AM 1.5 solar simulator system calibrated by standard silicon solar cell. The GO/rGO encapsulated and non-encapsulated devices were measured every 10 min for the first 30 min and every few hours afterwards; the devices are stored under room light in air (RH of 50-60 %) between each measurement.

**Acknowledgements**

The authors acknowledge K.Kuraoka of Kobe University, Japan and G.Eda of National University of Singapore for their technical supports at the initial stage of the work. Authors also acknowledge E.Cheng, J.Kim, and R.Kappera of Rutgers University for the experimental support, D.Watanabe of Tohoku University, Japan for the technical support. H.Y., D.V., M.C. acknowledge Donald H. Jacobs' Chair funding from Rutgers University. H.Y. acknowledges the Japanese Society for the Promotion of Science (JSPS) Postdoctoral Fellowship for Research Abroad, and Laboratory Directed Research and Development (LDRD) Director's Postdoctoral Fellowship of Los Alamos National Laboratory (LANL) for financial support. This research was funded in part by the Center on Materials and Devices for Information Technology Research (CDMITR), the National Science Foundation (NSF) grant #0120967, and Japan Science and Technology Agency (JST), PRESTO.

[1]    G. L. Graff, R. E. Williford, P. E. Burrows, *J. Appl. Phys.* **2004**, 96, 1840.
[2]    J. Meyer, P. Gorrn, F. Bertram, S. Hamwi, T. Winkler, H. H. Johannes, T. Weimann, P. Hinze, T. Riedl, W. Kowalsky, *Adv. Mater.* **2009**, 21, 1845.
[3]    J. Meyer, H. Schmidt, W. Kowalsky, T. Riedl, A. Kahn, *Appl. Phys. Lett.* **2010**, 96, 243308.
[4]    N. Kim, W. J. Potscavage, B. Domercq, B. Kippelen, S. Graham, *Appl. Phys. Lett.* **2009**, 163308.
[5]    J.S. Bunch, S.S. Verbridge, J.S. Alden, A.M. van der Zande, J.M. Parpia, H.G. Craighead, P.L. McEuen, *Nano Lett.* **2008**, 8, 2458.
[6]    A. Kolmakov, D.A. Dikin, L.J. Cote, J. Huang, M.K. Abyaneh, M. Amati, L. Gregoratti, S. Günther, M. Kiskinova, *Nat. Nanotech.*, **2011**, 6, 651.
[7]    S.P. Koenig, L. Wang, J. Pellegrino, J.S. Bunch, *Nat. Nanotech.*, **2012**, 7, 728.
[8]    K.S. Novoselov, V.I. Falko, L. Colombo, P.R. Gellert, M.G. Schwab, K. Kim, *Nature*, **2012**, 490, 192.
[9]    O. Leenaerts, B. Partoens, F.M. Peeters, *Appl. Phys. Lett.* **2008**, 93, 193105.
[10]   S. Chen, L. Brown, M. Levendorf, W. Cai, S.-Y. Ju, J. Edgeworth, X. Li, C.W. Magnuson, A. Velamakanni, R.D. Piner, J. Kang, J. Park, and R.S. Ruoff, *ACS Nano* **2011**, 1321.
[11]   M. Schriver, W. Regan, W.J. Gannett, A.M. Zaniewski, M.F. Crommie, A. Zettl, *ACS Nano* **2013**, DOI: 10.1021/nn4014356.
[12]   Z. Liu , J. Li , F. Yan, *Adv. Mater.* **2013**, DOI: 10.1002/adma.201205337.
[13]   W. S. Hummers and R. E. Offeman, *J. Am. Chem. Soc.* **1958**, 80, 1339.
[14]   M. Hirata, T. Gotou, S. Horiuchi, M. Fujiwara, M. Ohba, *Carbon* **2004**, 42, 2929.
[15]   S. Stankovich, D.A. Dikin, G.H.B. Dommett, K.M. Kohlhaas, E.J. Zimney, E.A. Stach, R.D. Piner, S.T. Nguyen, R.S. Ruoff,  *Nature* **2006**, 442, 282.
[16]   G. Eda, M. Chhowalla, *Adv. Mater.* **2010**, 22, 2392.

[17]     A. Bagri, C. Mattevi, M. Acik, Y.J. Chabal, M. Chhowalla, V.B. Shenoy, *Nat. Chem.* **2010**, 2, 581.
[18]     G. Eda, G. Fanchini, M. Chhowalla, *Nature Nanotech.* **2008**, 3, 270.
[19]     J.T. Robinson, M. Zalalutdinov, J.W. Baldwin, E.S. Snow, Z. Wei, P. Sheehan, B.H. Houston, *Nano Lett.* **2008**, 8, 3441.
[20]     H.Yamaguchi, G. Eda, C. Mattevi, H. Kim, M. Chhowalla, *ACS Nano* **2010**, 4, 524.
[21]     O.C. Compton, S. Kim, C. Pierre, J.M. Torkelson, S.T. Nguyen, *Adv. Mater.* **2010**, 22, 4759.
[22]     H. Kim, Y. Miura, C.W. Macosko, *Chem. Mat.* **2010**, 22, 3441.
[23]     Y.-H. Yang, L. Bolling, Morgan A. Priolo, J.C. Grunlan, *Adv. Mater.* **2013**, 25, 493.
[24]     M. Manceau, A. Rivaton, J.L. Gardette, S. Guillerez, N. Lemaitre, *Polym. Degrad. Stabil.* **2009**, 94, 898.
[25]     M. Manceau, S. Chambon, A. Rivaton, J. L. Gardette, S. Guillerez, N. Lemaitre, *Sol. Energ. Mat. Sol. C.* **2010**, 94, 1572.
[26]     L.J. Cote, R. Cruz-Silva, J. Huang, *J. Am. Chem. Soc.* **2009**, 131, 11027.
[27]     J. Granstrom, J.S. Swensen, J.S. Moon, G. Rowell, J. Yuen, A.J. Heeger, *Appl. Phys. Lett.* **2008**, 93, 193304.
[28]     P. Matyba, H. Yamaguchi, M. Chhowalla, N.D. Robinson, L.Edman, *ACS Nano* **2011**, 5, 574.
[29]     Cytop catalog, AGC Chemicals, Asahi Glass Co., Ltd., Tokyo, Japan, January, **2009**.
[30]     M. Acik, C. Mattevi, C. Gong, G. Lee, K. Cho, M. Chhowalla, Y.J. Chabal, *ACS Nano* **2010**, 4, 5861.
[31]     C. Mattevi, G. Eda, S. Agnoli, S. Miller, K.A. Mkhoyan, O. Celik, D. Mastrogiovanni, G. Granozzi, E. Garfunkel, M. Chhowalla, *Adv. Funct. Mater.* **2009**, 19, 1.
[32]     K .Erickson, R. Erni, Z. Lee, N. Alem, W. Gannett, A. Zettl, *Adv Mater.* **2010**, 22, 4467.
[33]     R.R. Nair, H.A. Wu, P.N. Jayaram, I.V. Grigorieva, A.K. Geim, *Science* **2012**, 335, 442.
[34]     C.J. Brabec, S. Gowrisanker, J.J.M. Halls, D. Laird, S. Jia, S.P. Williams, *Adv. Mater.* **2010**, 22, 3839.

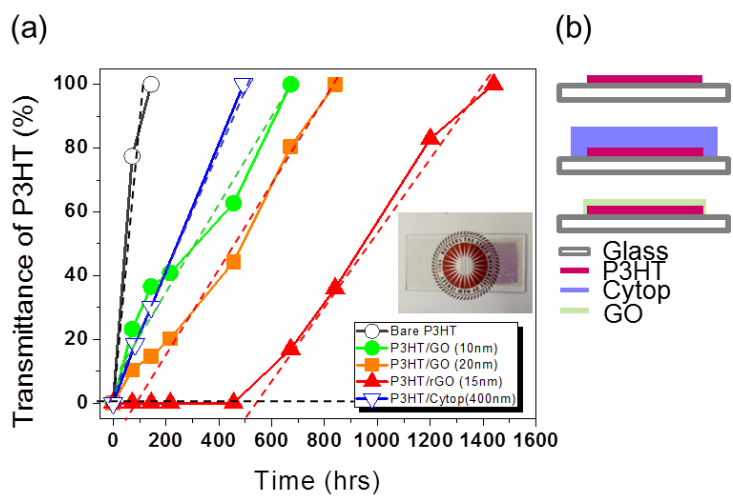

**Figure 1** Gas barrier property of GO/rGO films measured by P3HT transmittance test. **(a)** P3HT films on glass substrates were encapsulated with GO and rGO, and compared with controls. The inset shows the photograph of the GO films used for encapsulation on glass (left), and GO encapsulated P3HT films demonstrating high optical transparency of GO films used in this study (right). Dashed line fits used to extract the slope values for calculation of diffusivity in the steady-state flux regime are also shown. **(b)** Schematic of sample structures used for the P3HT transmittance tests.

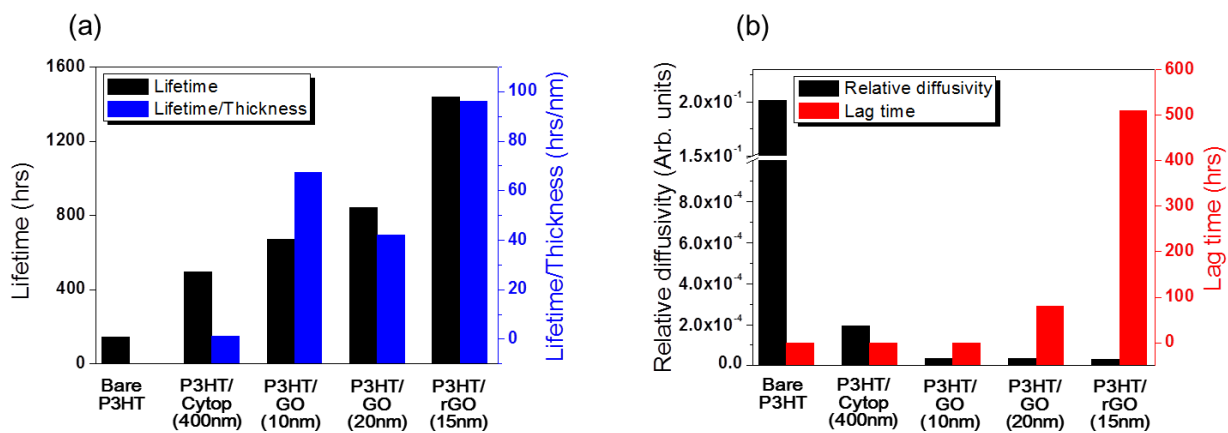

**Figure 2** Lifetime values of encapsulated P3HT films. **(a)** Lifetime values extracted from P3HT transmittance test shown for the various different barrier layers tested in this study. Black bars show the total lifetime in hours (left), and blue bars show the lifetime per encapsulated film thickness in hours/nm (right). **(b)** Permeation values showing the relative diffusivity of oxygen (black bars) and lag time for oxygen diffusion to occur (red bars).

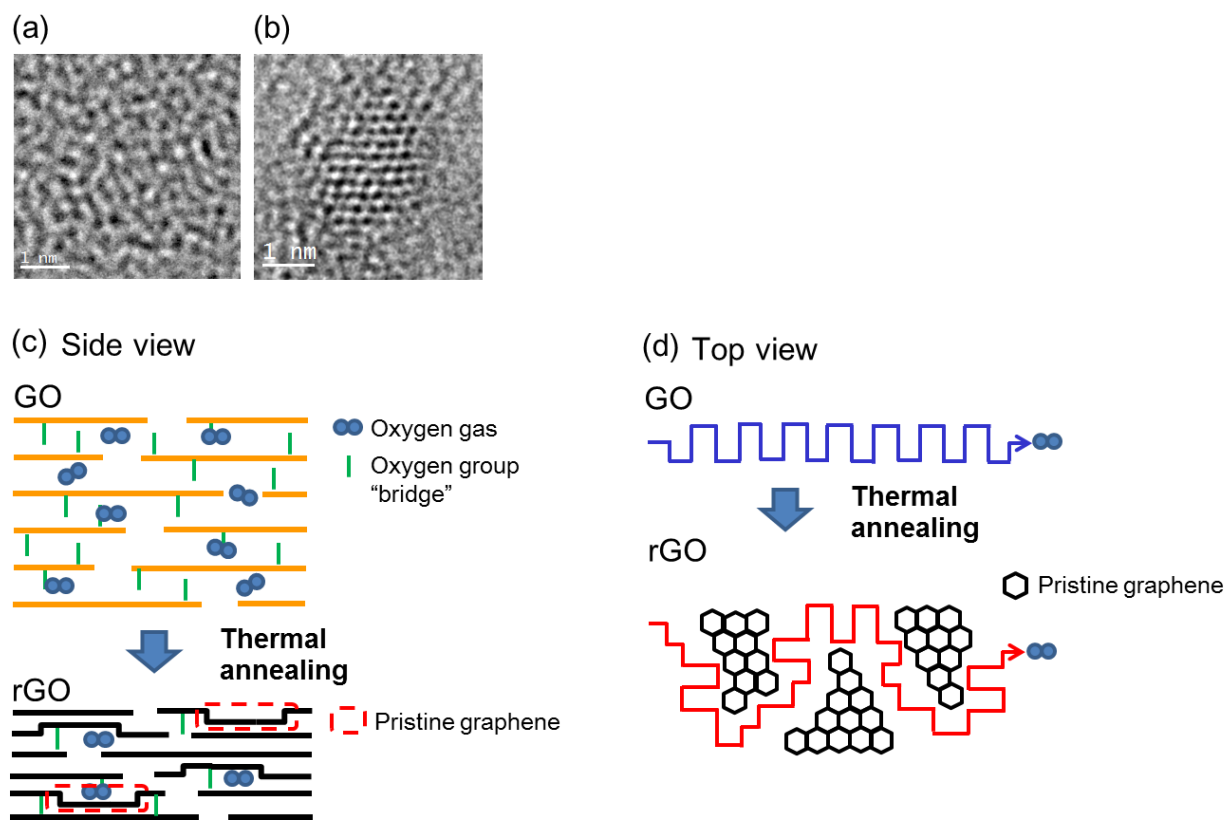

**Figure 3** Structural properties of GO/rGO barrier layers. High resolution transmission electron microscopy (HRTEM) image of GO **(a)** before and **(b)** after annealing. Restoration of hexagonal atomic structure in the form of graphene nano-island can be observed after annealing. Scale bars = 1 nm. **(c)** Schematics of side view showing oxygen gas permeating through GO (top) and rGO (bottom). For the case of GO, oxygen molecules (blue circles) permeates through the interlayer due to presence of oxygen functional groups acting as "bridges" (green bars). On the other hand, pristine graphene nano-islands created by removal of oxygen group "bridges" clogs permeation paths (dashed red boxes), increasing the diffusion length as illustrated in top view **(d)**. Possible permeation paths of the oxygen are indicated by the blue and red arrows for GO and rGO, respectively.

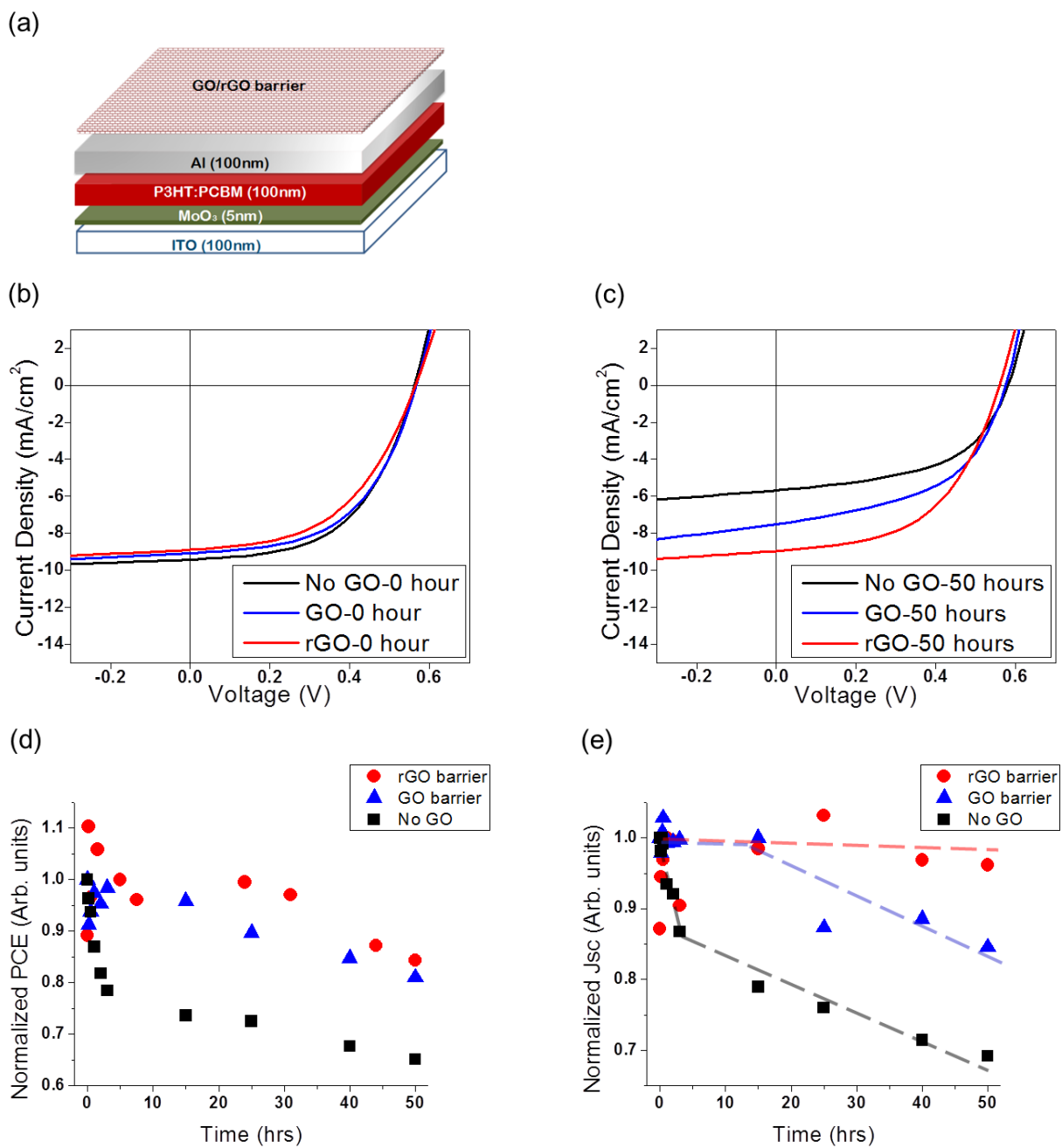

**Figure 4** rGO barriers for working organic photovoltaic devices. **(a)** Schematic of the tested device structure. ITO was on 1 mm thick glass substrate, and the illumination was exposed from bottom of the device through the glass substrate. Current-Voltage (I-V) characteristics of the devices for **(b)** 0 hour, and **(c)** 50 hours after exposure to ambient conditions. **(d)** Normalized power conversion efficiency (PCE), and **(e)** short circuit current ($J_{sc}$) *versus* tested time of GO/rGO encapsulated bulk heterojunction (BHJ) solar cells. Dashed lines are guide for the eye.

Supporting Information for

# Reduced graphene oxide thin films as ultrabarriers for organic electronics

*Hisato Yamaguchi\*, Jimmy Granstrom, Wanyi Nie, Hossein Sojoudi, Takeshi Fujita, Damien Voiry, Mingwei Chen, Gautam Gupta, Aditya D. Mohite, Samuel Graham, Manish Chhowalla\**

## S1 Two phases of gas permeation process through barrier films

Two dominant phases of gas permeation process through barrier films are i) transient, and ii) steady-state phase as shown in schematics below (**Figure S1**).[S1] In the transient phase, the permeation flux has not reached its equilibrium thus gas of interest is still in the process of permeating through the barrier film (red circle). In the steady-state phase, on the other hand, the permeation flux has reached its equilibrium giving constant flux rate. This results in linear relationship between total fluence and time as shown in the graph (blue circle). The constant flux rate of the gas in the steady-state phase can be extracted from the slope as diffusivity. The time obtained by explorating the slope of the steady-state phase is called lag time, which can be expressed by the equation (1) in the main text.

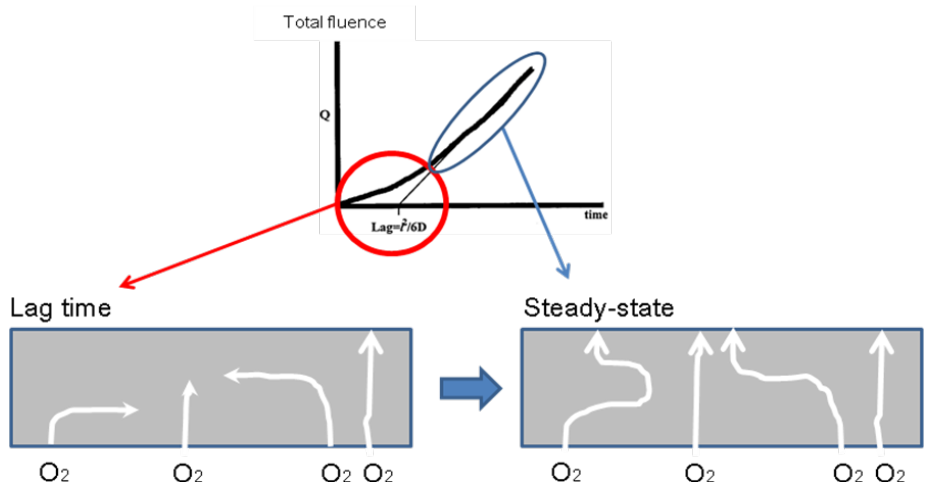

Figure S1

## S2 Diffusion length of GO barrier films

### S2.1 Flake size distribution of GO

**Figure S2.1** shows the flake size distribution of graphene oxide (GO) used in this study. Use of large starting graphite particles (>425 μm) for exfoliation, low rotation speed (5000 rpm) for centrifuge in the purification process, and eliminating ultrasonication from entire synthesis process led to distribution of relatively large GO flakes.

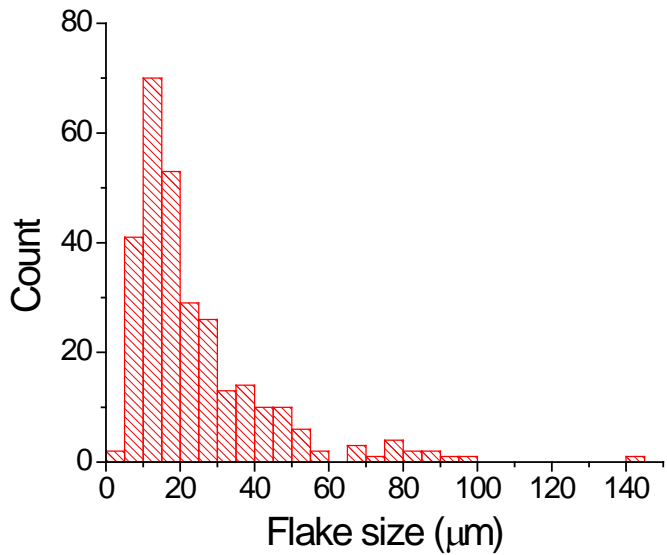

Figure S2.1

*S2.2 Estimated and experimentally obtained diffusion length of GO barrier film*

Using the average lateral GO flake size of 26.7 μm obtained from Fig. S2.1, rough estimation of the diffusion length results in 267 μm (0.267 mm) for the 20 nm thick film. Specifically, a diffusion length was estimated by multiplying half of the lateral sheet size (13.35 μm) by number of stacked layers, which is 20. The schematic of a permeation path used for the estimation is shown in **Figure S2.2**. This assumes that the gas permeates freely through the GO interlayer (in a straight manner). On the other hand, diffusion length was 0.365 cm (3.65 mm) based on equation (1) in the main text using the experimentally obtained lag time of 81.2 hrs ($2.92 \times 10^5$ s) and diffusivity of $7.6 \times 10^{-8}$ cm$^2$/s. The experimentally obtained diffusion length was factor of 14 larger than the simple estimation based on average lateral size and the number of GO layers. The difference between the two values strongly suggest that high density, and randomly distributed oxygen functional group "bridges" (Fig. 3 (a)) prevent gases to permeate freely (making gases to go round), causing an order of larger diffusion length compared to the straight paths shown in Fig. S2.2. The experimentally obtained results also supports that the oxygen permeating through the GO sheet in vertical direction is negligible, supporting our HRTEM observation of absence of pinholes.

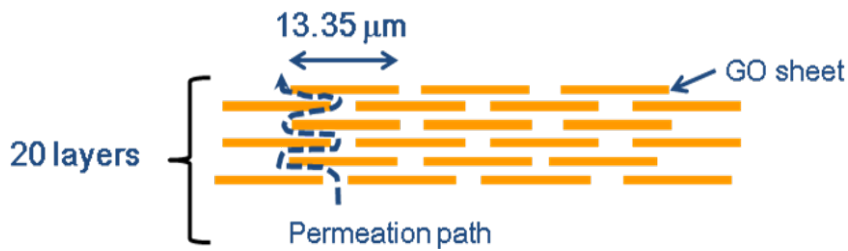

Figure S2.2

## S3 X-ray photoelectron spectroscopy (XPS) and atomic force microscopy (AFM) of GO and rGO

XPS and AFM were performed to investigate evolution of oxygen (O)/carbon (C) ratio and the thickness due to thermal reduction, respectively. **Figure S3** show the C1s peak of XPS spectra for GO films before and after thermal annealing. Annealing of GO films at 150 $^o$C resulted in evident decrease in the intensity of C-O binding energy peak at 286.5 eV, indicating that significant amount of oxygen functional groups were removed. The sharpening and slight intensity increase of the C-C peak at binding energy of 284 eV after the annealing was due to recovery of sp$^2$ carbon by removal of the oxygen groups. Quantitative analysis using O/C ratio indicated that the oxygen decreased from 54 at.% to 37 at.%, consistent with our previous study.[S2] The removal of oxygen functional groups (and water) due to annealing was also consistent with decrease of film thicknesses characterized by AFM. The thickness of the thick GO films decreased from 20 nm to 15 nm after annealing at 150 $^o$C, which is in good agreement with the literature.[S3] Detailed analysis results of XPS C1s peak showing atomic % for four dominating oxygen functional groups (C-OH: hydroxyl, C-O: epoxy, C=O: carbonyl, and COO: carboxyl) as well as sp$^2$ bonding of carbon atoms before and after annealing at 150 °C are shown in Table S3. Above mentioend oxgygen functional groups attract moisture and affect the degradation process of the encapsulated P3HT films/organic electronic devices.

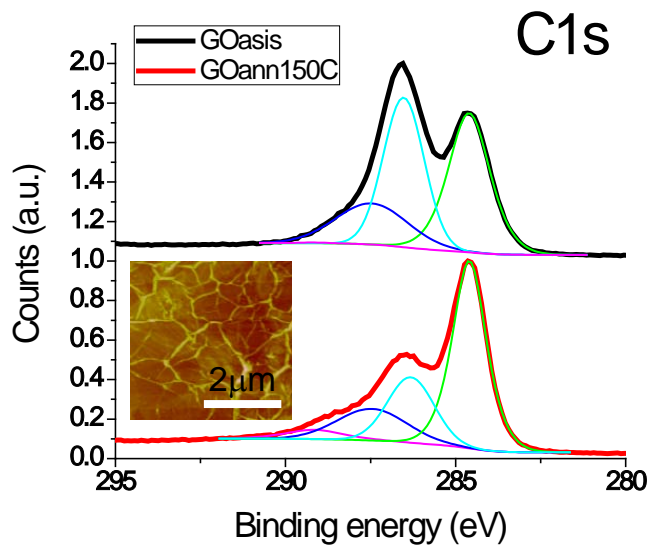

Figure S3 Evolution of XPS C1s spectra for GO films annealed at 150 °C. Each color of deconvoluted peak indicates the following chemical bonds. Green: C-C, Light blue: C-O, Blue: C=O, Pink: COO. Inset shows a typical AFM image for thick annealed GO films. Note that there are randomly distributed wrinkles which are the origin of larger ratio thickness increase per layer for thicker films. The density of wrinkles was much lower for thin GO films. Scale bar is 2 μm.

| Temperature (°C) | C-C(sp$^2$) | C-OH | C-O | C=O | COO |
|---|---|---|---|---|---|
| RT | 44.3 | 15.2 | 13.3 | 5.2 | 3.5 |
| 150 | 54.9 | 11.4 | 9.4 | 3.2 | 1.7 |

Table S3 XPS C1s peak analysis results showing atomic % for four dominating oxygen functional groups and sp$^2$ bonding of carbon atoms for before and after annealing at 150 °C.

## *S4 Absolute power conversion efficiency (PCE) of GO thin film encapsulated bulk heterojunction (BHJ) solar cells*

**Figure S4** shows the absolute power conversion efficiency (PCE)-time characteristics of GO thin film encapsulated bulk heterojunction (BHJ) solar cells. The initial absolute PCE of BHJ solar cells used in this study ranged between 3.6-3.2 %.

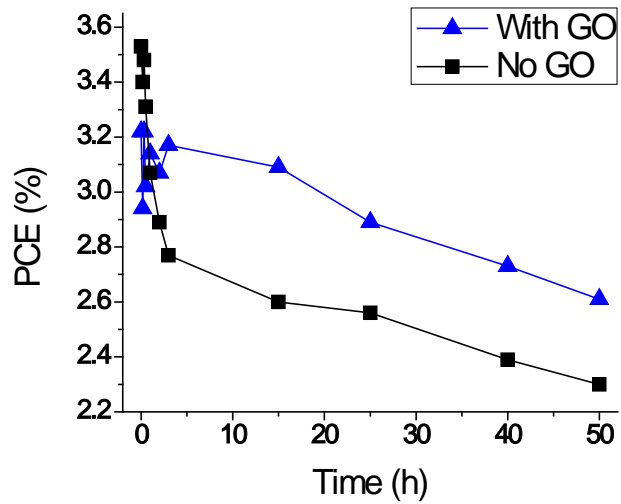

Figure S4

**Table S4** shows changes of absolute open circuit voltage (Voc), short circuit current (Jsc), and fill factor (FF), and PCE for the unencapsulated BHJ solar cells left in air at room temperature for one day (24 hours). The results indicated that the drastic drop of PCE from 3.33 % to 0.48 % was strongly correlated with the decrease of Jsc, while Voc and FF remained almost constant. The decrease of Jsc suggests Jsc as a parameter that is most sensitive to oxygen diffusion into the devices, which is known as a primary degradation factor for the PCE.

|  | Voc (V) | Jsc (mA/cm^2) | FF | PCE (%) |
|---|---|---|---|---|
| Fresh | 0.603 | 9.20 | 60 | 3.33 |
| One day | 0.603 | 1.51 | 53.3 | 0.48 |

Table S4

## S5 Raman spectroscopy of rGO encapsulated P3HT films

Raman spectroscopy of rGO encapsulated P3HT films was performed to demonstrate that the transmittance test results are correlated to the structural changes of P3HT (**Figure S5** (a)). Fig. S5 (b) shows the results for unencapsulated case as a comparison. The same test conditions were used as the transmittance tests except for the higher illumination intensity to accelerate the degradation process of P3HT. Specifically, 100 W/cm$^2$ AM 1.5 solar simulator light source was used for the Raman lifetime tests instead of 0.22 W/cm$^2$ used for the transmittance lifetime tests. Raman spectroscopy was performed on a micro Raman spectrometer (InVia Raman microscope, Renishaw) with a 785 nm excitation laser. The laser power has been reduced to ~5 mW to avoid any damage to the samples. The Raman shift peak of our interest was at 1440 cm$^{-1}$ as indicated in red arrows of Fig. S5 (a), (b). The peak is the fingerprint of P3HT assigned to C=C ring stretching, a bond which is known to be crucial for high performance P3HT based electronic devices.[S4, S5] The results clearly showed that the structures of P3HT were preserved after rGO thin film encapsulation, giving qualitatively same trend as the transmittance test results (Fig. S5 (c)).

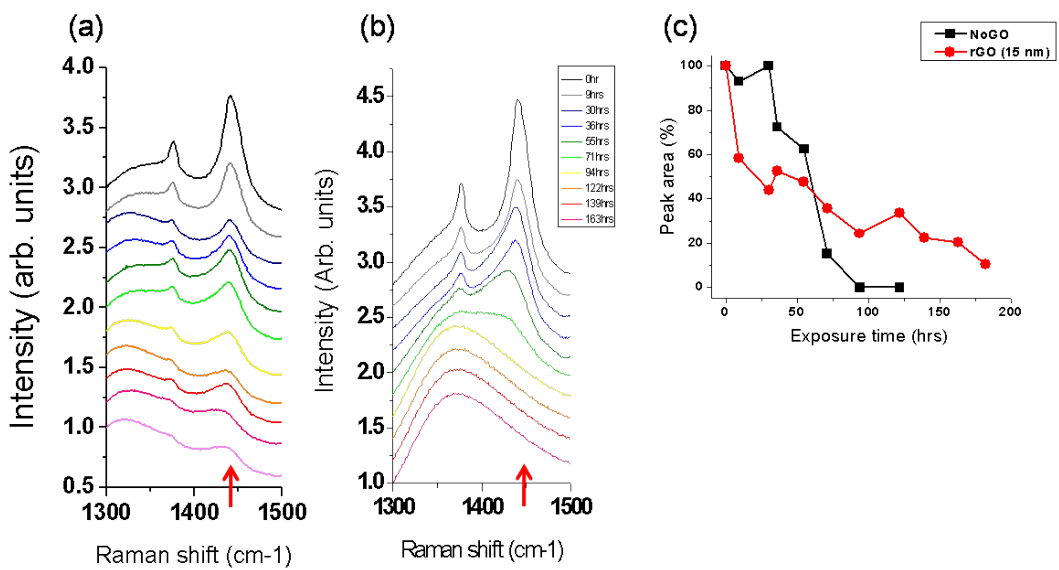

Figure S5 (a) Raman spectroscopy results of rGO encapsulated P3HT films under illumination in ambient condition. (b) Without encapsulation as a control. Red arrows show P3HT fingerprint peak at 1440 cm$^{-1}$ assigned to C=C ring stretching. (c) Peak area percentage of P3HT fingerprint peak at 1440 cm$^{-1}$ verses illumination exposure time for rGO encapsulated (red) and unencapsulated (black) films.

[S1] G.L. Graff, R.E. Williford, P.E. Burrows, *J. Appl. Phys.* **2004**, 96, 1840.
[S2] C. Mattevi, G. Eda, S. Agnoli, S. Miller, K.A. Mkhoyan, O. Celik, D. Mastrogiovanni, G. Granozzi, E. Garfunkel, M. Chhowalla, *Adv. Funct. Mater.* **2009**, 19, 1.
[S3] I. Jung, M. Vaupel, M. Pelton, R. Piner, D.A. Dikin, S. Stankovich, J. An, R.S. Ruoff, *J. Phys. Chem. C*, **2008**, 112, 8499.
[S4] S. Miller, G. Fanchini, Y.-Y. Lin, C. Li, C.-W. Chen, W.-F. Su, M. Chhowalla, *J. Mater. Chem.* **2008**, 18, 306.
[S5] W.C. Tsoi, D.T. James, J.S. Kim, P.G. Nicholson, C.E. Murphy, D.D.C. Bradley, J. Nelson, J.-S. Kim, *J. Am. Chem. Soc.* **2011**, 133, 9834.